\documentclass[preprint,notoc]{JHEP3}
\usepackage{graphics,slashed}

\newcommand{\be}{\begin{equation}} 
\newcommand{\ee}{\end{equation}} 
\newcommand{\bea}{\begin{eqnarray}} 
\newcommand{\eea}{\end{eqnarray}}

\def\spa#1.#2{\langle#1\,#2\rangle}
\def\spb#1.#2{[#1\,#2]}
\def\sandmm#1.#2.#3{%
\left\langle\smash{#1}{\vphantom1}\right|{#2}%
\left|\smash{#3}{\vphantom1}\right]}
\def\spab#1.#2.#3{\sandmm#1.#2.#3}
\def\spba#1.#2.#3{\sandpp#1.#2.#3}
\def\spaa#1.#2.#3.#4{\sandmp#1.{#2#3}.#4}
\def\spbb#1.#2.#3.#4{\sandpm#1.{#2#3}.#4}
\def\spash#1.#2{\spa{\smash{#1}}.{\smash{#2}}}
\def\spbsh#1.#2{\spb{\smash{#1}}.{\smash{#2}}}

\def\ksl{\not{\hbox{\kern-2.3pt $k$}}}

\def\e{\epsilon}

\def\Ord{{\cal O}}

\def\NeqFour{{\cal N}=4}

\def\li#1{{\mathop{\rm Li}\nolimits}_#1}

\preprint{RUNHETC-2009-23}

\title{The Imaginary Part of the $\NeqFour$ Super-Yang-Mills \\ \hspace{.25 in}Two-Loop Six-Point MHV Amplitude in \\ \hspace{1 in} Multi-Regge Kinematics}

\author{Robert M. Schabinger\\
	Department of Physics and Astronomy\\
	Rutgers, The State University of New Jersey\\
      136 Frelinghuysen Rd. Piscataway, NJ 08854}

\abstract{The precise form of the multi-Regge asymptotics of the two-loop six-point MHV amplitude in $\NeqFour$ Super-Yang-Mills theory has been a subject of recent controversy. In this paper we utilize the amplitude/Wilson loop correspondence to obtain precise numerical results for the imaginary part of these asymptotics. The region of phase-space that we consider is interesting because it allowed Bartels, Lipatov, and Sabio Vera to determine that the two-loop six-point MHV amplitude is not fixed by the BDS ansatz. They proceeded by working in the framework of a high energy effective action, thus side-stepping the need for an arduous two-loop calculation. Our numerical results are consistent with the predictions of Bartels, Lipatov, and Sabio Vera for the leading-log asymptotics.}


\dedicated{\centerline{Submitted to JHEP}}

\begin{document}

%%%%%%%%%%%%%%%%%%%%%%%%%%%%%%%%%%%%%%%%%%%%%%%%%%%

\section{Introduction}\label{IntroSection}

In recent years, tremendous progress has been made towards a more complete understanding of the gluon scattering amplitudes in $\NeqFour$ Super-Yang-Mills theory (hereafter simply $\NeqFour$). Particularly interesting is the remarkable fact that, in the planar limit, it is possible to completely solve the perturbative S-matrix (up to momentum independent pieces) for the scattering of either four gluons or five gluons. Building on the earlier work of~\cite{ABDK}, Bern, Dixon, and Smirnov (BDS) made an all-loop, all-multiplicity proposal for the finite part of the maximally-helicity-violating (MHV) amplitudes in $\NeqFour$. In this paper~\cite{BDS}, BDS explicitly demonstrated that their ansatz was valid for the four-point amplitude through three loops. Subsequent work demonstrated that the BDS ansatz holds for the five-point amplitude through two-loops~\cite{TwoLoopFive} and that the strong coupling form of the four-point amplitude has precisely the form predicted by BDS~\cite{AM1}.  

In fact,~\cite{AM1} sparked a significant parallel development. Motivated by the fact that the strong coupling calculation proceeded by relating the four-point gluon amplitude to a particular four-sided light-like Wilson loop, the authors of~\cite{DHKSWL4} were able to show that the finite part of the four-point light-like Wilson loop at weak coupling matches the finite part of the four-gluon scattering amplitude. The focus of~\cite{DHKSWL4} was on the four-gluon MHV amplitude, but it was shown in~\cite{BHTWLn} that this MHV amplitude/light-like Wilson loop correspondence holds for all one-loop MHV amplitudes  in $\NeqFour$. It is clear that an arbitrary $n$-gluon light-like Wilson loop should be conformally invariant in position space.\footnote{Strictly speaking, the conformal symmetry is anomalous due to the presence of divergences at the cusps in the Wilson loop. If one regulates these divergences and subtracts the conformal anomaly, then the finite part of what remains will be conformally invariant.} What was not at all obvious before the discovery of the amplitude/Wilson loop correspondence is that $\NeqFour$ scattering amplitudes must be (dual) conformally invariant in {\it momentum} space. 

It turns out that this hidden symmetry (referred to hereafter as dual conformal invariance) has non-trivial consequences for the $\NeqFour$ S-matrix. Assuming that the MHV amplitude/light-like Wilson loop correspondence holds to all loop orders, the authors of~\cite{DHKSWL2l4} were able to prove that dual conformal invariance  fixes the (non-perturbative) form of all the four- and five- point gluon helicity amplitudes (non-MHV amplitudes first enter at the six-point level) in $\NeqFour$. Up to trivial factors, they showed that the functional form of the (dual) conformal anomaly coincides with that of the BDS ansatz. Subsequently, work has been done~\cite{BM,BRTW} that provides some evidence for the assumption made in~\cite{DHKSWL2l4} that the MHV amplitude/light-like Wilson loop correspondence holds to all orders in perturbation theory.  

The idea is that, due to the fact that non-trivial conformal cross-ratios can first be formed at the six-point level, one would naively expect the four- and five-point amplitudes to be momentum-independent constants to all orders in perturbation theory. It is well-known, however, that gluon loop amplitudes have IR divergences. These IR divergences explicitly break the dual conformal symmetry and it is precisely this breaking which allows four- and five- gluon loop amplitudes to have non-trivial momentum dependence. In fact, the arguments of~\cite{DHKSWL2l4} allowed the authors to predict the precise form that the answer should take and they found (up to trivial constants) complete agreement with the BDS ansatz to all orders in perturbation theory.

At this stage, it was unclear whether the appropriate hexagon Wilson loop would still be dual to the six-point MHV amplitude at the two-loop level. This question was decisively settled in the affirmative by the work of~\cite{BDKRSVV} on the scattering amplitude side and~\cite{DHKSWL2l6b} on the Wilson loop side. Another issue settled by the authors of~\cite{BDKRSVV} and~\cite{DHKSWL2l6b} was the question of whether the BDS ansatz fails at two loops and six points. It had already been pointed out by Alday and Maldacena in~\cite{AM3} that the BDS ansatz must fail to describe the analytic form of the $L$-loop $n$-gluon MHV amplitude for sufficiently large $L$ and $n$, but it had not yet been conclusively proven until the appearance of~\cite{BDKRSVV} and~\cite{DHKSWL2l6b} that $L = 2$ and $n = 6$ was the simplest possible example of BDS ansatz violation.

In point of fact, if one was only interested in the question of BDS ansatz violation, it would not have been necessary to perform the complete two-loop calculation. It was already known to Bartels, Lipatov, and Sabio Vera (BLSV) from computations they performed~\cite{BLSV1} that the BDS ansatz fails at two-loops and six-gluons. It is these computations that we will scrutinize in the present paper. For simplicity, we restrict ourselves to the regime of multi-Regge kinematics (MRK regime), where all gluons are strongly ordered in rapidity. The argument of ~\cite{BLSV1} relied only on the general structure of the multi-Regge limit; it was in~\cite{BLSV2} that the explicit formula for the imaginary part of the analytically continued\footnote{We postpone a detailed discussion of the phase-space slice under consideration until the next section.} $2 \rightarrow 4$ amplitude in the MRK regime was actually presented. 

The goal of the current project is to take the Feynman integral representation of the $\NeqFour$ two-loop six-point MHV amplitude as given\footnote{As a practical matter, the Feynman integrals appearing in the Wilson loop calculation of~\cite{DHKSWL2l6b} seem to be much easier to evaluate numerically than the Feynman integrals appearing in the scattering amplitude calculation of~\cite{BDKRSVV}.} by~\cite{DHKSWL2l6b} and analytically continue (before numerical evaluation) all integrals in the same way that the authors of~\cite{BLSV2} analytically continued their representation of the amplitude. It is then possible to make a direct comparision between exact numerical results and results which should describe the leading-log behavior as one goes deep into the MRK regime. The main result of this paper is that, when appropriately interpreted, we find solid agreement between our  numerical results for the imaginary part and the predictions of the approximate formula derived by BLSV.

In our view this undertaking was necessary because the authors of~\cite{V}, Del Duca, Duhr, and Glover (DDG), have expressed doubts as to whether BLSV correctly identified the physics of the situation in question. Normally, one expands the amplitude in a Laurent series in the dimensional regularization parameter $\e$ (defined as $d = 4 - 2\e$) before performing any analytical continuations of the amplitude. DDG claim that one can only obtain a BLSV-like formula if one goes to the MRK regime and performs all desired analytical continuations before expanding the scattering amplitude in $\e$. To the author's knowledge,~\cite{H} is the only other article that has considered the results of BLSV and DDG.\footnote{In the first version of this article remarks were made about~\cite{H} that have been removed in this version. The reason for this change is that, in going from v3 of~\cite{H} to the final published version, the discussion of BLSV and DDG changed significantly. We thank Howard Schnitzer for bringing this to our attention.} In section 8 of~\cite{H}, the authors of that work make it clear that they believe the methods of~\cite{BLSV2} are correct. Their paper, however, focuses on somewhat different issues and does not address the question quantitatively. Our goal is to demonstrate conclusively that there are no problems with the work of BLSV (at least for the $2 \rightarrow 4$ amplitude that we have consider in detail) and it is hoped that our results will put to rest the controversy surrounding the results presented in~\cite{BLSV2}.

This article is organized as follows.  
In section~\ref{prelim} we establish our notation and describe in more detail the problem which we have solved.  In section~\ref{body} we present the main results of this paper and compare them to those of BLSV. In section~\ref{ConclusionSection} we present our conclusions, explain how our work sheds light on some claims made in reference~\cite{V}, and discuss some open problems appropriate for future work. In Appendix \ref{KSection} we review six-body kinematics and the relation between scattering amplitude and Wilson loop notation. Finally, in Appendix~\ref{ACSection}, we give some formulas useful for analytical continuation and an explicit example of one of the more intricate calculations we performed. 

%%%%%%%%%%%%%%%%%%%%%%%%%%%%%

\section{Preliminaries}
\label{prelim}

As mentioned in the introduction, we do not deal with the two-loop six-gluon MHV amplitude directly. Instead, we use the two-loop light-like hexagon Wilson loop which is known to be equal to the amplitude of interest~\cite{BDKRSVV,DHKSWL2l6b} in a sense that will be made clear below. Recall that, for a given multiplicity, there is effectively only one MHV amplitude in $\NeqFour$, due to the power of supersymmetric Ward identities (see reference~\cite{CVthesis} for a review of these ideas). This fact suggests that it is most natural to consider MHV loop amplitudes divided by the appropriate Parke-Taylor~\cite{PT} tree amplitude. The way in which one formally expresses the correspondence between the finite parts of 2-loop six-gluon helicity amplitudes and the 2-loop light-like Wilson loop is
\be
\ln \Bigg({A^{\NeqFour}_{6}(1^-,2^-,3^+,4^+,5^+,6^+)\over A^{\rm tree}_{6} (1^-,2^-,3^+,4^+,5^+,6^+)}\Bigg)\Bigg|_{{\rm finite};~\Ord(a^2)}  \sim \ln \bigg(\langle A^{\rm light-like~ WL}_{6}\rangle \bigg)\Bigg|_{{\rm finite};~\Ord(a^2)} \, ,
\label{equiv}
\ee
where we have chosen the $(1^-,2^-,3^+,4^+,5^+,6^+)$ helicity configuration for sake of definiteness\footnote{Note that this does not imply an equivalence between the $\Ord(a^2)$ finite parts of $A^{\NeqFour}_{6}(1^-,2^-,3^+,4^+,5^+,6^+)$ divided by the tree amplitude and $\langle A^{\rm light-like~ WL}_{6} \rangle$. See references~\cite{7and8WL} and~\cite{Cachazo} for details.}. What we really mean by the above equivalence is that, if the logs on either side are expanded to $\Ord(a^2)$ (the 't Hooft coupling, $a$, is defined below), the finite parts of the $\Ord(a^2)$ coefficients should match up to an additive constant independent of the dynamics. The additive constant which turns the above equivalence into an equality  was given in~\cite{7and8WL}.\footnote{It should also be pointed out that the $\Ord(\e^{-1})$ pole terms do not exactly match. The physics of this mismatch was properly understood in~\cite{DMS}.}

Before we try to make sense of Eq.~(\ref{equiv}), let us first remind ourselves of the overall normalization used for the numerical evaluation of the finite part of the hexagon Wilson loop.  In~\cite{DHKSWL2l6b}, the exact color structure was computed and only at the end of the calculation did the authors take the $N_c \rightarrow \infty$ limit. They were then able to pull an overall factor of $C_F N_c ( g^2 / 4 \pi^2)^2$  out of the finite part of all Feynman diagrams. Finally, they made the replacement $C_F N_c (g^2 / 4 \pi^2)^2 \rightarrow 2 a^2$. This replacement is that which is required to obtain numerical results consistent with the definition used by BLSV for the 't Hooft coupling:
\be
a = {g^2 N_c\over 8 \pi^2} \, .
\ee

As mentioned above, proceeding naively results in a mismatch between the two-loop hexagon Wilson loop and the two-loop six-point MHV amplitude. To understand this problem, we need a little more information about the form of the expected answer. On the amplitude side, the answer is constrained by dual conformal invariance to look like
\be
\ln \Bigg({A^{\NeqFour}_{6}(1^-,2^-,3^+,4^+,5^+,6^+)\over A^{\rm tree}_{6} (1^-,2^-,3^+,4^+,5^+,6^+)}\Bigg)\Bigg|_{{\rm finite};~\Ord(a^2)} = F_{6;2}^{{\rm BDS}} + R_{6;2}(u_1,u_2,u_3) \, ,
\label{2Lstruct}
\ee
where $F_{6;2}^{{\rm BDS}}$ is the finite part predicted by BDS in~\cite{BDS} (we use the conventions of~\cite{DHKSWL2l6b} for this function) and the $u_i$ are conformal cross-ratios built out of the standard cyclicly symmetric Mandelstam invariants (see Appendix \ref{KSection} for a discussion of the essential points):
\be
u_1 = {s_{1} s_{4}\over t_{1}t_{3}}~~~~~~~~~~u_2 = {s_{2} s_{5}\over t_{1}t_{2}}~~~~~~~~~~~~u_3 = {s_{3} s_{6}\over t_{2}t_{3}} \, .
\ee
$R_{6;2}(u_1,u_2,u_3)$ is simply called the (two-loop six-point) remainder function. By construction, $R_{6;2}(u_1,u_2,u_3) \rightarrow 0$ in any collinear limit. Operationally, this means that $R_{6;2}(u_1,u_2,u_3) \rightarrow 0$ as $u_i \rightarrow 1;~u_{i+1},~u_{i+2} \rightarrow 0$ ($i$ mod 3). 

However, if one uses the dictionary between Mandelstam invariants in the scattering amplitude picture and non-zero lengths in the Wilson loop picture (Appendix \ref{KSection}), one finds that, on the Wilson loop side, the answer looks like~\cite{7and8WL}
\be
\ln \bigg(\langle A^{\rm light-like~ WL}_{6}\rangle \bigg)\Bigg|_{{\rm finite};~\Ord(a^2)} = F_{6;2}^{{\rm BDS}} + \bigg(R_{6;2}(u_1,u_2,u_3)+{\pi^4 \over 8}\bigg) \, .
\ee
Therefore, before we can utilize the Wilson loop picture calculations of~\cite{DHKSWL2l6b}, we have to adjust their formulas by subtracting an overall factor of $\pi^4/8$ on the right-hand side. This gurantees that the Wilson loop version of $R_{6;2}(u_1,u_2,u_3)$ has the correct (vanishing) collinear limit.

We now describe precisely what the MRK regime is in the context of this paper. Typically, when one performs a loop-level gauge theory calculation one takes all invariants to be negative and only analytically continues the result back into the physical region after all Feynman integrals have been evaluated\footnote{Four-body scattering is a good example. Physically, one is typically interested in a $2 \rightarrow 2$ process with $s>0$ and $t<0$. On the other hand, for the evaluation of Feynman integrals, it is convenient to take all particles outgoing and both $s$ and $t$ negative. }. The region of phase-space with all invariants negative (hereafter referred to as the Euclidean region) is special because all Feynman integrals evaluate to real numbers. We will eventually be interested in a specific analytical continuation, but for now we discuss the Euclidean MRK regime.

The Euclidean MRK regime is one in which all of the Mandelstam invariants are space-like and their absolute values satisfy the following ordering relations
\be
s_1 >> s_5, s_4, s_3 >> s_6, t_2, s_2 \,\,\,,
\label{ordering}
\ee
where we have suppressed the absolute value bars for the sake of simplicity. Any phase-space slice that is conformally equivalent to one satisfying (\ref{ordering}) is suitable for studying the remainder function in the MRK regime. In fact, for this purpose one can define the MRK regime completely in terms of $u_1$, $u_2$, and $u_3$. In this regime all of the particles are collinear (either left or right movers) and, consequently, two of the conformal cross-ratios approach zero (from above) and one of them approaches unity (from below). Which cross-ratio approaches unity depends on how one labels the external momenta. The particular labeling scheme that we have made in this work coincides with that of~\cite{BLSV2} and gives $u_1 \rightarrow 1$ and~$u_{2},~u_{3} \rightarrow 0$. In any case, the fact that the remainder function is unchanged by cyclic permutations of the cross-ratios~\cite{DHKSWL2l6b} ensures that all choices lead to the same physics. We refer the interested reader to the introduction sections of~\cite{V},~\cite{H}, and~\cite{BNST1} for a much more detailed discussion. As expected, we find that the remainder function in the Euclidean MRK regime is approaching zero. Due to the fact that the remainder function approaches zero in the Euclidean MRK regime, it is not suitable for detecting BDS ansatz violation at the two-loop six-point level.

Instead we consider a particular analytical continuation of the Euclidean MRK regime that preserves the conformal cross-ratios. The analytic continuation is one where both Mandelstam invariants in the numerator of $u_1$, the cross-ratio approaching unity, are allowed to change sign. The nice thing about such an analytical continuation is that one can set up the problem in the Euclidean MRK regime and then analytically continue by simply making the replacements $-s_1 \rightarrow s_1$ and $-s_4 \rightarrow s_4$. BLSV claim that, even though $R_{6;2}(u_1,u_2,u_3) \rightarrow 0$ in the Euclidean MRK regime, the remainder function develops a non-vanishing imaginary part (that grows logarithmically) if one performs the analytical continuation described above, thus providing a slick way to check that the BDS ansatz fails at the two-loop six-point level. We evaluate this claim quantitatively in the next section. 
\section{The Imaginary Part}
\label{body}

In principle there is no need to restrict oneself to the calculation of the imaginary part, but we do so both because this is what BLSV presented in~\cite{BLSV2} and because the imaginary part turns out to be much easier to control numerically. To make the calculation as simple as possible, we attempted to parametrize the non-vanishing distances in a way that makes the number of Feynman integrals that need to be analytically continued as small as possible. Unfortuately, some complication is unavoidable, since several of the higher-dimensional Feynman integrals have to be processed further in the Euclidean region (beyond what was done in the Appendix of~\cite{DHKSWL2l6b}) to put them in a form suitable for analytical continuation. Specifically, we found that, at least for numerical evaluation in {\tt Mathematica}, all Feynman diagrams had to be expressed in terms of integrals which were two-fold or less. Before continuing, we first made sure that all of the contributing integrals were correctly entered into {\tt Mathematica} by reproducing Table 1 of~\cite{DHKSWL2l6b}.

The primary phase-space slice that we worked on can be parametrized in terms of the three positive real numbers $a$, $b$, and $c$ and is given by
\bea
x_{13}^2 = -c ;\;\
x_{24}^2 = -a ;&& \;\
x_{35}^2 = -b ;\;\
x_{46}^2 = -c ;\;\
x_{15}^2 = -a ;\;\
x_{26}^2 = -b \nonumber
\\
x_{14}^2 &&= -1 ;\;\
x_{25}^2 = -1 ;\;\
x_{36}^2 = -1 \, .
\label{slice1}
\eea
On this slice we have the cross-ratios $u_1 = c^2$, $u_2 = a^2$, and $u_3 = b^2$, where 
\be
c = a+b-1+2\sqrt{1-a-b+ a b}
\label{gramdeterm}
\ee
is fixed by the Gram determinant constraint (see Appendix~\ref{KSection} for details).

Analytical continuation to the region of interest is performed by replacing $-c$ with $c$ and then carefully following the rules outlined in Appendix \ref{ACSection}. We found that, apart from several trivial factors similar to those that appear in the BDS ansatz, we had to continue 84 integrals away from the Euclidean region to extract the complete imaginary part of $R_{6;2}(c^2,a^2,b^2)$ on this particular branch.

Carrying out the calculation is quite arduous, due to the fact that {\tt Mathematica} is not equipped to implement the rules described in Appendix \ref{ACSection} automatically. Fortunately, conformal invariance can be used as a stringent cross-check on the final results. One has to look at a different kinematic point that has the same conformal cross-ratios and make sure that $R_{6;2}(c^2,a^2,b^2)$ is defined on the same branch after the analytical continuation ($-c \rightarrow c$) has been performed. We generated a secondary test configuration along these lines given by 
\bea
x_{13}^2 = -c ;\;\
x_{24}^2 = -a ;&& \;\
x_{35}^2 = -{2 b \over 3} ;\;\
x_{46}^2 = -3 c ;\;\
x_{15}^2 = -a ;\;\
x_{26}^2 = -2 b \nonumber
\\
x_{14}^2 &&= -{3 \over 2} ;\;\
x_{25}^2 = -{2 \over 3} ;\;\
x_{36}^2 = -2 \, .
\label{slice2}
\eea
It is easily checked that this configuration preserves the cross-ratios. Analytically continuing this configuration in the manner described above, we confirmed that the result for the imaginary part of $R_{6;2}(c^2,a^2,b^2)$ on (\ref{slice2}) is equal to the imaginary part of $R_{6;2}(c^2,a^2,b^2)$ on (\ref{slice1}) to within the available numerical precision.

Once we established confidence that our {\tt Mathematica} script for the primary configuration was correct, we extrapolated our results deeper into the MRK regime and compared them to the explicit formula given by BLSV~\cite{BLSV2} 
\be
{\rm Im} \Big\{ R_{6;2}(u_1,u_2,u_3) \Big\} = {\pi \over 2}\ln\big({1 - u_1}\big)\ln\Big({1 - u_1 \over u_1~ u_2}\Big)\ln\Big({1 - u_1 \over u_1~ u_3}\Big) +  \cdots
\ee
in terms of dual conformal cross-ratios (we've pulled out two factors of $a$ to make contact with the conventions of ~\cite{DHKSWL2l6b}). The dots represent sub-leading-log terms not determined by the analysis in~\cite{BLSV2}. 

At this stage the skeptical reader may be wondering how we intend to make a meaningful statement about the BLSV function using numerical analysis when the sub-leading corrections to it are completely unknown. There are really two issues that need to be addressed: Does the BLSV function capture the leading-log behavior of the imaginary part? If so, how much do the sub-leading corrections matter? In general, it is not guranteed that numerical analysis can be used to address these questions. Fortunately, the problem at hand has a very special structure which allows us to make progress.

Before discussing our results, we have to explain how we parametrized the problem. We were motivated by the structure of the Gram determinant constraint to think about expanding the expression for $u_1(u_2,u_3)$ in $(u_2,u_3)$ about $(0,0)$. This allows us to Taylor expand Eq.~(\ref{gramdeterm})
\be
1-u_1 \approx {\Big(\sqrt{u_2}-\sqrt{u_3}\Big)^2 \over 2}
\label{Taylor}
\ee
and drop the higher-order terms. In this approximation, we can rewrite the BLSV formula in terms of just two variables, which we choose to be $u_2$ and $\xi = u_2/u_3$:
\be
{\rm Im} \Big\{ R_{6;2}(u_2,\xi) \Big\} = {\pi \over 2}\ln\Bigg(u_2 {\Big(1-{1\over \sqrt{\xi}}\Big)^2 \over 2}\Bigg)\ln\Bigg({\Big(1-{1\over \sqrt{\xi}}\Big)^2 \over 2}\Bigg)\ln\Bigg({\Big(1-\sqrt{\xi}\Big)^2 \over 2}\Bigg) +  \cdots \, .
\label{BLSVform2}
\ee
Note that Eq.~(\ref{BLSVform2}) depends on $u_2$ through the first logarithm only. This means that on lines of constant $\xi$ the BLSV function looks like
\be
A \ln(u_2)+B+\cdots 
\label{BLSVform3}
\ee
with $A$ and $B$ given by 
\bea
A &=& {\pi \over 2}\ln\Bigg({\Big(1-{1\over \sqrt{\xi}}\Big)^2 \over 2}\Bigg)\ln\Bigg({\Big(1-\sqrt{\xi}\Big)^2 \over 2}\Bigg) \label{A}\\
B &=& {\pi \over 2}\Bigg[\ln\Bigg({\Big(1-{1\over \sqrt{\xi}}\Big)^2 \over 2}\Bigg)\Bigg]^2 \ln\Bigg({\Big(1-\sqrt{\xi}\Big)^2 \over 2}\Bigg)
\label{B}
\eea
plus the unknown subleading-log contributions which one might think could depend on $u_2$ and $u_3$ in a complicated way, even for $u_2$ and $u_3$ very close to zero.  

In fact, it turns out that the sub-leading terms depend only on $\xi$ to leading order. If the sub-leading logs depend only on $\xi$ (a fact that we justify {\it a posteriori}), then grouping $B(\xi)$ with the $A(\xi) \ln(u_2)$ term in the BLSV function is an arbitrary choice; for the purposes of this analysis we could equally well group it with the unknown corrections because it too depends only on $\xi$. Moreover, the assumption that the unknown contributions depend only $\xi$ suggests a simple experiment to see whether the BLSV formula actually does capture the leading-log behavior of the imaginary part. For a given $\xi$, we take nine data points evenly spaced on a log scale ($6.9444 \times 10^{-7} \leq u_2 \leq 0.0016$), differentiate\footnote{To differentiate our data with respect to $\ln(u_2)$ we took two data points very close together and computed $u_2 {\Delta {\rm Im} \{ R_{6;2}(c^2, ~a^2,~ b^2) \}\over \Delta u_2}$ for the pair. We then experimented with different seperations in $u_2$ to make sure that the results were stable.} with respect to $\ln(u_2)$ and then plot the results (Figure \ref{figure1}) against Eq.~(\ref{A}) as a function of $\ln(u_2)$. The range of $\ln(u_2)$ plotted was fixed for us by the breakdown of our numerics for very small $u_2$ and the breakdown of our approximation scheme (Eq.~(\ref{Taylor}))  for relatively large $u_2$. We performed the analysis for four evenly spaced values of the ratio $\xi$ on the unit interval. Recall that the imaginary part is symmetric under the interchange of $u_2$ and $u_3$. This implies that we are free to fix $u_3 > u_2$ without loss of generality. In this case, $(0,1)$ is the natural domain for $\xi$.\footnote{At $u_3 = u_2$ there is a branch point and for sufficiently large $u_3$ (sufficiently small $\xi$) we will necessarily leave the MRK regime if we consider contours of constant $\xi$ over the range of $\ln(u_2)$ plotted in Figure \ref{figure1}.} 

We find that the numbers extracted from our data match those from Eq.~(\ref{A}) very well indeed. The fact that this works at all justifies our assumption that the unknown sub-leading-log contributions depend only on $\xi$ to first order in the small quantities $u_2$ and $u_3$. Furthermore, it shows that our exact results are consistent with a non-trivial prediction of BLSV as a function of $\xi$. It should also be stressed that, although, we produced values of $u_1$ by solving the Gram determinant constraint in the Euclidean MRK regime for our primary kinematic configuration, this was not a necessary step and we could have obtained qualitatively identical results without the Gram determinant constraint. After all, our secondary test configuration did not satisfy the Gram determinant constraint and yet, by virtue of the conformal invariance, it matched a configuration that {\it did} satisfy the constraint. For example, if we scaled $1-u_1$ to zero at {\it exactly} the same rate that $u_2$ and $u_3$ were scaled to zero we would have been able to produce plots qualitatively equivalent to those displayed in Figure \ref{figure1}. 

\FIGURE{
\resizebox{.8\textwidth}{!}{\includegraphics{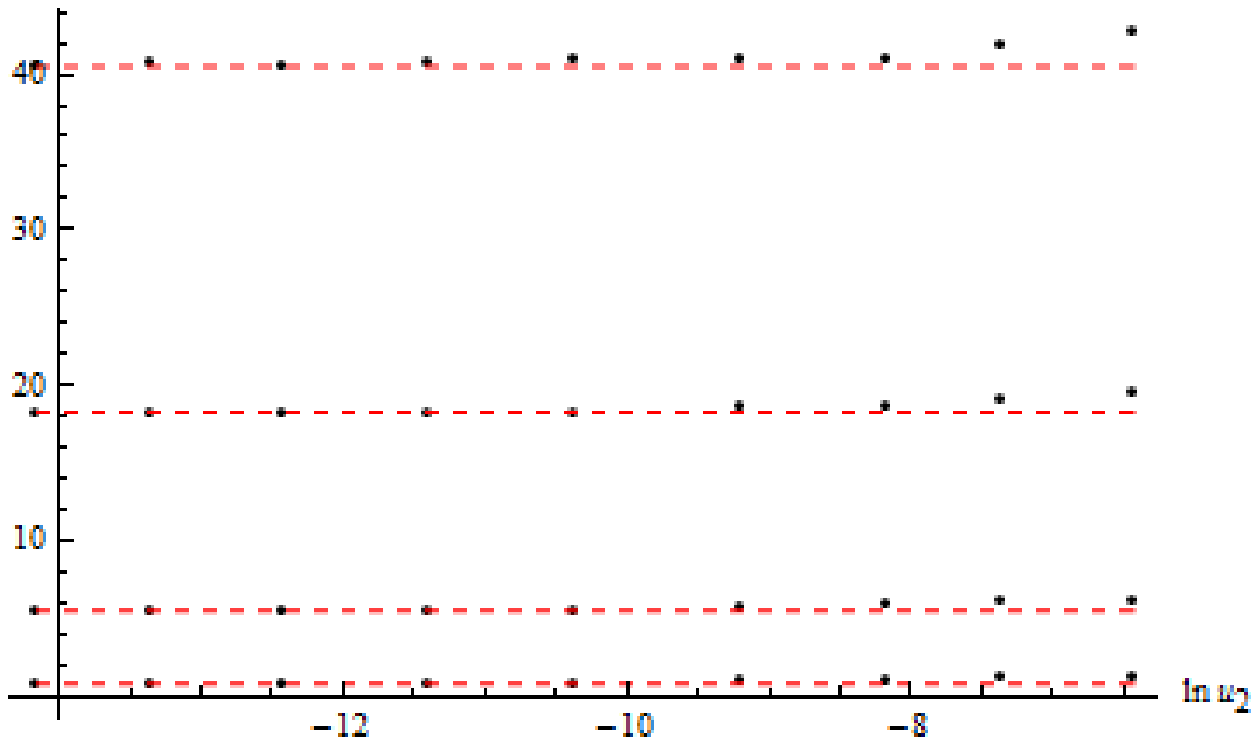}}
\caption{The derivative of our data with respect to $\ln(u_2)$ (black dots) and the coefficient $A$ of Eq.~(\ref{A}) (red dashed) are plotted as a function of $\ln(u_2)$. From bottom to top, results for $u_2/u_3$ ($\xi$) ratios 0.2, 0.35, 0.6, and 0.8 are shown. Note that the lowest-order approximation used for $1-u_1$ (Eq.~(\ref{Taylor})) starts to break down for $\ln(u_2)$ greater than about -8.}
\label{figure1}}

Of course, there is a piece of Eq.~(\ref{BLSVform2}) (the function $B(\xi)$ given by Eq.~(\ref{B})) that cannot be detected by taking derivatives with respect to $\ln(u_2)$. We expect that it is present by virtue of the fact that the analytical continuation performed in~\cite{BLSV2} was in the invariants $s_1$ and $s_4$ sitting inside of $u_1$, but we have no way of nailing it down. Although we can't say much about $B(\xi)$ or the unknown $\xi$ dependent corrections, it would be nice to get a qualitative feel for how close the BLSV function comes to describing the complete imaginary part. Of course, the phase-space slice considered above does not allow one to do this because two of the three logarithms in the formula were nearly invariant on it. We can do better by varying $\xi$ keeping $u_2$ fixed. We plotted our data for $u_2 = 1.5625\times 10^{-6}$ (an arbitrarily chosen point that was found to be numerically stable and solidly within the regime of validity of the small $(u_2,u_3)$ approximation) and varied $u_3$ from 1.0609 $u_2$ to 36 $u_2$. These results are compared to the BLSV function and $A(\xi)\ln(u_2)$ from Eq.~(\ref{BLSVform3}) in Figure \ref{figure2}. We see that the general shape of the curves is predicted pretty well by BLSV but that the sub-leading logs become progressively more important as the branch point at $\xi = 1$ is approached. The fact that $A(\xi)\ln(u_2)$ (the blue dot-dashed curve) fits the data plotted in Figure \ref{figure2} better than the actual BLSV formula (red dashed curve) suggests that the true leading-log behavior is described by $A(\xi)\ln(u_2)$ on the phase-space slice under consideration (Eq.~(\ref{slice1})) and that it is more natural to group $B(\xi)$ with the undetermined sub-leading-log contributions.

\FIGURE{
\resizebox{.8\textwidth}{!}{\includegraphics{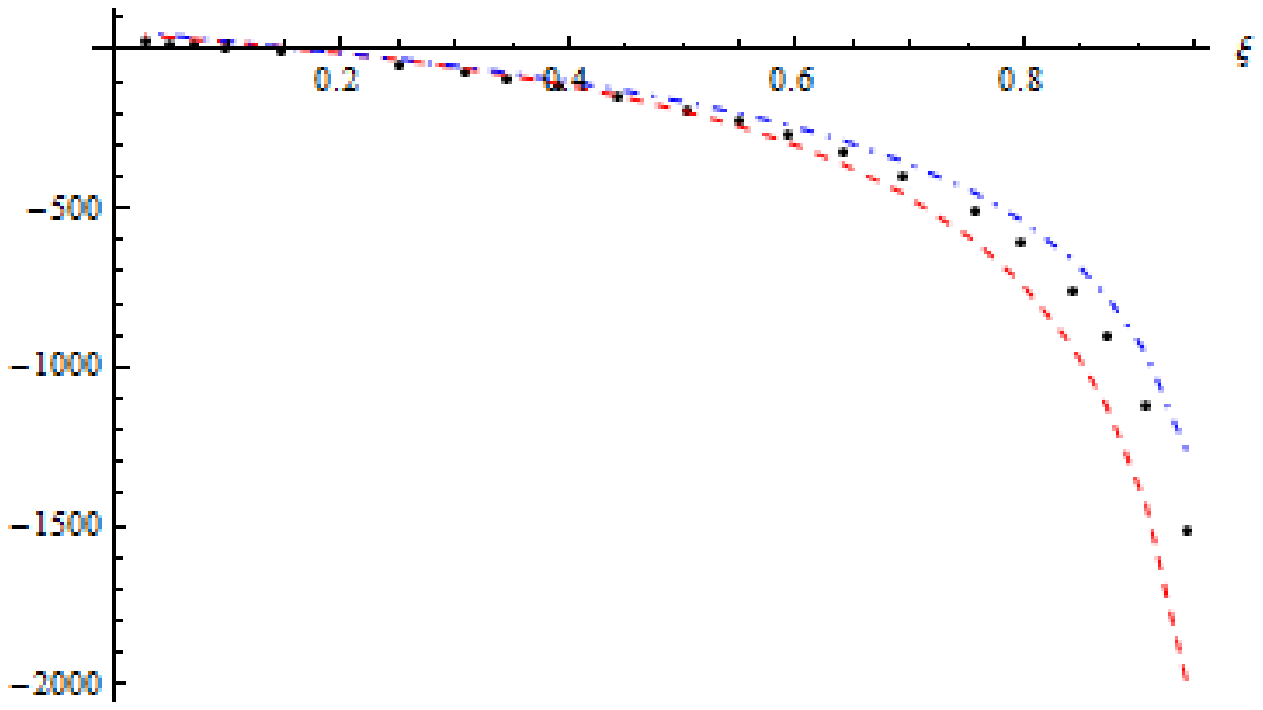}}
\caption{Our data (black dots), the BLSV function (red dashed), and $A(\xi) \ln(u_2)$ from Eq.~(\ref{BLSVform3}) (blue dot-dashed) are plotted as a function of $u_2/u_3$ ($\xi$) for $u_2 = 1.5625 \times 10^{-6}$. Presumably, the significant deviations present for large $\xi$ (for both curves) are due to the presence of the undetermined sub-leading-log contributions to the imaginary part.}
\label{figure2}}

\section{Discussion and Conclusions}
\label{ConclusionSection}
In this work we studied the claim made by BLSV that a particular analytical continuation of the Euclidean MRK regime allows one to see that the BDS ansatz fails for the two-loop six-gluon MHV amplitude in $\NeqFour$. We find that our precise numerical results for the imaginary part of the amplitude are completely consistent with the formula derived by BLSV in~\cite{BLSV2}. Our work provides strong evidence that the interpretation of the physics given in the work of BLSV is correct, at least for the $2 \rightarrow 4$ process that we considered in detail. Our work also bears on a couple of issues raised by DDG.

In~\cite{V}, DDG claim that it is not possible to see the imaginary part reported by BLSV unless one analytically continues away from the Euclidean MRK regime {\it before} expanding the amplitude in $\e$. We encountered no such difficulty. In other words, we have no reason to believe that there is any problem with the conventional approach to perturbative gauge theory calculations in this context. Furthermore, DDG remark that the kinematical regime chosen by BLSV is unphysical. It is not clear to us what DDG mean by this statement. In any case, we show in Appendix \ref{KSection} that, if desired, it is possible to solve the Gram determinant constraint {\it after} analytical continuation ($-c \rightarrow c$) on the phase-space slice defined by Eq.~(\ref{slice1}). 

Finally, we highlight a point of potential confusion, also discussed in~\cite{H}. One might worry that the analytical continuation described in Section \ref{body} is ill-defined because the solutions to the Gram determinant constraint on a phase-space slice like the one defined in Eq.~(\ref{slice1}) are not invariant under $-c \rightarrow c$ (compare \ref{sol1} and \ref{sol2} to \ref{sol3}). Our analysis has explicitly shown that a well-defined imaginary part exists regardless of whether the Gram determinant constraint is satisfied. This is most clearly seen by recalling that we were able to obtain results using (\ref{slice2}) identical to those found by using (\ref{slice1}), even though the kinematic configuration defined by (\ref{slice2}) did not satisfy the Gram determinant constraint before or after the analytic continuations were performed. Several authors~\cite{DHKSWL2l6b,7and8WL} have already concluded that the (dual) conformal invariance of the remainder function is independent of the Gram determinant constraint. Although their calculations were performed in the Euclidean region, our work has shown explicitly  that the imaginary part of the remainder function behaves in a completely analogous way. 

Our main goal in this endeavor was to bring some clarity to the existing literature. One deficiency of the present paper is that the analysis we have presented is purely numerical and is therefore somewhat limited in scope. It may prove useful to find an analytical formula for the imaginary part we have studied, because this result could yield useful information about the functional form of the remainder function for general kinematics. It is also quite likely that this line of attack will be simpler than tackling the general problem directly because the space of transcendentality three functions is smaller than the space of transcendentality four functions.

\section*{Acknowledgments}

First and foremost I am deeply indebted to Lance Dixon for suggesting this line of research and for many stimulating discussions. I would also like to thank the SLAC theory group for its hospitality the last few summers. I am also very grateful to James Drummond, Johannes Henn, Gregory Korchemsky, and Emery Sokatchev for their
enlightening correspondence and for sharing 
details of their work with me. Last but certainly not least, I would like to thank Matt Strassler for his continued support and guidance.
This research was supported by the US Department of 
Energy under contract DE-FG02-96ER40949.
\appendix

\section{Six-Body Kinematics and the Amplitude/Wilson Loop Correspondence}
\label{KSection}
In this appendix, we review some facts about six-body kinematics and establish a dictionary that makes the mapping between amplitudes and Wilson loops concrete. 

Although there are only eight independent Mandelstam invariants in a six-body scattering process~\cite{phspints}, it is more convenient for our purposes to use a set of nine variables that arise naturally in the propagator denominators of Feynman diagrams contributing to the scattering process. Taking all external momenta to be outgoing, these variables are defined by the relations
\be
s_{i} = \big(k_i+k_{i+1}\big)^2 \hspace{.25 in} \textrm{and} \hspace{.25 in} t_{j} = \big(k_j+k_{j+1}+k_{j+2}\big)^2 \, ,
\ee
where the index $i$ is taken mod six and then index $j$ is taken mod three. This basis of Mandelstam invariants has the nice feature that it forms a cyclicly symmetric set.

There are several references in the body of this paper to the ``Gram determinant constraint''. This constraint is what makes only eight of the nine variables defined above independent quantities. The constraint is simply expressed~\cite{Dim1loop} as
\be
{\rm Det'}(2 k_i \cdot k_j) = 0 \, ,
\ee
where the prime signifies that one of the six four-vectors in the problem is to be omitted in taking the determinant. To be concrete we omit $k_6$. In this case, we have
\bea
 &&(2 k_i \cdot k_j)_{k_6} = \\  &&\left(\begin{array}{ccccc} 0 & s_1 & t_1-s_1-s_2 & s_2+s_5-t_1-t_2 & t_2-s_5-s_6  \\ s_1 & 0 & s_2 & t_2-s_2-s_3 & s_3+s_6-t_2-t_3  \\ t_1-s_1-s_2 & s_2 & 0 & s_3 & t_3-s_3-s_4  \\ s_2+s_5-t_1-t_2 & t_2-s_2-s_3 & s_3 & 0 & s_4  \\ t_2-s_5-s_6 & s_3+s_6-t_2-t_3 & t_3-s_3-s_4 & s_4 & 0 \end{array}\right)\,. \nonumber
 \label{Sij}
\eea
There are actually two ways that one could set ${\rm Det}(2 k_i \cdot k_j)_{k_6}$ equal to zero for the kinematics defined by  Eq.~(\ref{slice1}) and still be in the Euclidean MRK regime. Although we chose the solution 
\be
c = a + b - 1 + 2\sqrt{1-a-b+a b}
\label{sol1}
\ee
for the analysis performed in this paper, we equally well could have chosen
\be
c = \sqrt{1-a^2-b^2+a^2 b^2} - a b
\label{sol2}
\ee
because both (\ref{sol1}) and (\ref{sol2}) force $c$ to approach 1 from below when $a$ and $b$ are taken to zero from above. It is also worth noting that, if desired, one could solve the Gram determinant constraint in the MRK regime {\it after} analytical continuation ($-c \rightarrow c$). In this case, there is a unique consistent choice given by
\be
c = \sqrt{1-a^2-b^2+a^2 b^2} + a b \, .
\label{sol3}
\ee
We intentionally worked with kinematics where the Gram determinant constraint was satisfied in the Euclidean MRK regime before analytical continuation ($c(a, b)$ defined by Eq.~(\ref{sol1})) because we wanted to see explicitly that it does not affect the analysis if the Gram determinant constraint fails to be satisfied after the replacement $-c \rightarrow c$ has been made. 

Finally, the mapping between distances in the Wilson loop picture and the canonical basis of Mandelstam variables defined above is expressed as
\bea
x_{13}^2 &&\leftrightarrow ~~s_{1} \nonumber
\\
x_{24}^2 &&\leftrightarrow ~~s_{2} \nonumber
\\
x_{35}^2 &&\leftrightarrow ~~s_{3} \nonumber
\\
x_{46}^2 &&\leftrightarrow ~~s_{4} \nonumber
\\
x_{15}^2 &&\leftrightarrow ~~s_{5} \nonumber
\\
x_{26}^2 &&\leftrightarrow ~~s_{6} \nonumber
\\
x_{14}^2 &&\leftrightarrow ~~t_{1} \nonumber
\\
x_{25}^2 &&\leftrightarrow ~~t_{2} \nonumber
\\
x_{36}^2 &&\leftrightarrow ~~t_{3}
\eea
\section{Analytical Continuation}
\label{ACSection}
In this appendix, we discuss the essentials of analytical continuation as it pertains to the results presented in this paper. As an example, we give a complete derivation of the imaginary part for the integrals associated to topology $(j)$ in~\cite{DHKSWL2l6b}. All of the formulae below are easily derived from the results given in Appendix C of~\cite{AC}. Here we simply give a representative sample of the results needed for the particular problem studied in this paper. In what follows, $s$ and $t$ should be thought of as negative real numbers such that $s t < 1$. 
\bea
\ln\bigg({s \over -t}\bigg) &&= \ln\bigg({-s \over -t}\bigg) - i \pi \label{form1}\\
\ln\bigg({-s \over t}\bigg) &&= \ln\bigg({-s \over -t}\bigg) + i \pi \label{form2}\\
\ln\bigg({s \over t}\bigg) &&= \ln\bigg({-s \over -t}\bigg) \label{form3}\\
\li2\bigg(1-{s \over -t}\bigg) &&= {\pi^2 \over 3}-{1\over2} \ln^2\bigg(1+{-s \over -t}\bigg)-\li2\bigg({1 \over 1+{-s \over -t}}\bigg) + i \pi \ln\bigg(1+{-s \over -t}\bigg) \\
\li2\bigg(1-{-s \over t}\bigg) &&= {\pi^2 \over 3}-{1\over2} \ln^2\bigg(1+{-s \over -t}\bigg)-\li2\bigg({1 \over 1+{-s \over -t}}\bigg) - i \pi \ln\bigg(1+{-s \over -t}\bigg) \\
\li2\bigg(1-{s \over t}\bigg) &&= \li2\bigg(1-{-s \over -t}\bigg)\\
\li2\bigg(1-s t\bigg) &&= \li2\bigg(1-(-s)(-t)\bigg) + 2 i \pi \ln\bigg(1-(-s)(-t)\bigg)\\
\li2\bigg(1-{1 \over s t}\bigg) &&= \li2\bigg(1-{1\over(-s)(-t)}\bigg) - 2 i \pi \ln\bigg(1-{1\over s t}\bigg) \\
&&= \li2\bigg(1-{1\over(-s)(-t)}\bigg) + 2  \pi^2 - 2 \pi i \ln\bigg({1\over (-s) (-t)}-1\bigg)
\eea

The starting point for topology $(j)$ is Eq.~(A.23) of~\cite{DHKSWL2l6b}. Though we technically have a Wilson loop integral, we express everything in terms of $s_i$ and $t_j$ using the dictionary of Appendix \ref{KSection} because it makes the formulas easier to read. It is also convenient to define
\be
N(s_1,s_4,t_1,t_3,v_1,v_2) = t_3-(t_3-s_4)v_1-(t_3-s_1)v_2-(s_1+s_4-t_1-t_3)v_1 v_2.
\ee
After a little reorganization, we pull out two factors of the 't Hooft coupling and sum over distinguishable cyclic permutations to obtain
\bea
&&{1\over4}\int_0^1 d v_1\int_0^1 d v_2 {1\over {t_3-s_1 \over s_1 + s_4 - t_1 - t_3}+v_1}\ln\bigg({N(s_1,s_4,t_1,t_3,v_1,v_2)\over s_1 (1 - v_1) + t_1 v_1}\bigg) \times \nonumber \\
&&\times {1 \over {t_3-s_4 \over s_1 + s_4 - t_1 - t_3}+v_2}\ln\bigg({N(s_1,s_4,t_1,t_3,v_1,v_2)\over t_3 (1 - v_2) + s_1 v_2}\bigg) + \nonumber \\
&&{1\over4}\int_0^1 d v_1\int_0^1 d v_2 {1\over {t_1-s_2 \over s_2 + s_5 - t_2 - t_1}+v_1}\ln\bigg({N(s_2,s_5,t_2,t_1,v_1,v_2)\over s_2 (1 - v_1) + t_2 v_1}\bigg) \times \nonumber \\
&&\times {1 \over {t_1-s_5 \over s_2 + s_5 - t_2 - t_1}+v_2}\ln\bigg({N(s_2,s_5,t_2,t_1,v_1,v_2)\over t_1 (1 - v_2) + s_2 v_2}\bigg) + \nonumber \\
&&{1\over4}\int_0^1 d v_1\int_0^1 d v_2 {1\over {t_2-s_3 \over s_3 + s_6 - t_3 - t_2}+v_1}\ln\bigg({N(s_3,s_6,t_3,t_2,v_1,v_2)\over s_3 (1 - v_1) + t_3 v_1}\bigg) \times \nonumber \\
&&\times {1 \over {t_2-s_6 \over s_3 + s_6 - t_3 - t_2}+v_2}\ln\bigg({N(s_3,s_6,t_3,t_2,v_1,v_2)\over t_2 (1 - v_2) + s_3 v_2}\bigg) \, .
\eea
On the slice defined by Eq.~(\ref{slice1}), this becomes
\bea
&&\int_0^1 d v_1\int_0^1 d v_2 {\ln\bigg({N(-c,-c,-1,-1,v_1,v_2)\over c(1-v_1)+v_1}\bigg) \ln\bigg({N(-c,-c,-1,-1,v_1,v_2)\over 1-v_2+c v_2}\bigg) \over (1-2 v_1)(1-2 v_2)} + \nonumber \\
&&\int_0^1 d v_1\int_0^1 d v_2 {\ln\bigg({N(-a,-a,-1,-1,v_1,v_2)\over a(1-v_1)+v_1}\bigg) \ln\bigg({N(-a,-a,-1,-1,v_1,v_2)\over 1-v_2+a v_2}\bigg) \over (1-2 v_1)(1-2 v_2)} + \nonumber \\
&&\int_0^1 d v_1\int_0^1 d v_2 {\ln\bigg({N(-b,-b,-1,-1,v_1,v_2)\over b(1-v_1)+v_1}\bigg) \ln\bigg({N(-b,-b,-1,-1,v_1,v_2)\over 1-v_2+b v_2}\bigg) \over (1-2 v_1)(1-2 v_2)}  \, .
\eea
Clearly, only the first of these terms develops an imaginary part under $-c \rightarrow c$. After dropping the other terms and making this substitution, we find that we are faced with analytically continuing
\be
\int_0^1 d v_1\int_0^1 d v_2 {\ln\bigg({N(c,c,-1,-1,v_1,v_2)\over -c(1-v_1)+v_1}\bigg) \ln\bigg({N(c,c,-1,-1,v_1,v_2)\over 1-v_2-c v_2}\bigg) \over (1-2 v_1)(1-2 v_2)}
\label{topj}
\ee
on the unit square in the $v_1-v_2$ plane. A glance at Eqs. (\ref{form1}), (\ref{form2}), and (\ref{form3}) tells us that we must isolate the regions of this parameter space on which either the numerator or denominator (but not both simultaneously) of either logarithm change sign. The conditions for the denominators to change sign are simple because they are expressed in terms of constants. For the denominator of the first logarithm the condition is
\be
v_1 \leq {c\over 1+c}
\ee
and for the denominator of the second logarithm the condition is
\be
v_2 \geq {1\over 1+c} \, .
\ee
The condition for the numerator is more complicated. We choose to express the conditions as a function of $v_2$ and they are then given by the inequalities
\bea
v_1 \geq {1 - v_2 (1+c)\over (1+c)(1-2 v_2)} ~~~~~{\rm if} ~~~~~v_2 \leq {c \over 1+c} \nonumber \\
v_1 \leq {1 - v_2 (1+c)\over (1+c)(1-2 v_2)} ~~~~~{\rm if} ~~~~~v_2 \geq {1 \over 1+c} \, .
\eea
It isn't possible for the numerators to change sign if $v_2$ is between $c/(1+c)$ and $1/(1+c)$. 

We now systematically investigate how the numerators and denominators of the logarithms in (\ref{topj}) change sign as one moves between different domains. Let us begin by parametrizing the problem. In $v_2$ there are three regimes: $0 \leq v_2 \leq c/(1+c)$, $c/(1+c) \leq v_2 \leq 1/(1+c)$, and $1/(1+c) \leq v_2 \leq 1$. In the first, we find three different regions: $0 \leq v_1 \leq c/(1+c)$, $c/(1+c) \leq v_1 \leq (1 - v_2 (1+c))/(1+c)/(1-2 v_2)$, and $(1 - v_2 (1+c))/(1+c)/(1-2 v_2) \leq v_1 \leq 1$. In the second, we find just two: $0 \leq v_1 \leq c/(1+c)$ and $c/(1+c) \leq v_1 \leq 1$. Finally, in the third regime, we again find three separate regions: $0 \leq v_1 \leq (1 - v_2 (1+c))/(1+c)/(1-2 v_2)$, $(1 - v_2 (1+c))/(1+c)/(1-2 v_2) \leq v_1 \leq c/(1+c)$, and $c/(1+c) \leq v_1 \leq 1$.

We now describe what happens in each of these eight regions. If $0 \leq v_2 \leq c/(1+c)$ and $0 \leq v_1 \leq c/(1+c)$, then the denominator of the first logarithm in (\ref{topj}) changes sign. If $0 \leq v_2 \leq c/(1+c)$ and $c/(1+c) \leq v_1 \leq (1 - v_2 (1+c))/(1+c)/(1-2 v_2)$, then nothing happens. If $0 \leq v_2 \leq c/(1+c)$ and $(1 - v_2 (1+c))/(1+c)/(1-2 v_2) \leq v_1 \leq 1$, then both numerators change sign. If $c/(1+c) \leq v_2 \leq 1/(1+c)$ and $0 \leq v_1 \leq c/(1+c)$, then the denominator of the first logarithm changes sign. If $c/(1+c) \leq v_2 \leq 1/(1+c)$ and $c/(1+c) \leq v_1 \leq 1$, then nothing happens. If $1/(1+c) \leq v_2 \leq 1$ and $0 \leq v_1 \leq (1 - v_2 (1+c))/(1+c)/(1-2 v_2)$, then all numerators and denominators change sign, effectively the same as if nothing at all had happened. If $1/(1+c) \leq v_2 \leq 1$ and $(1 - v_2 (1+c))/(1+c)/(1-2 v_2) \leq v_1 \leq c/(1+c)$, then both denominators change sign. Finally, if $1/(1+c) \leq v_2 \leq 1$ and $c/(1+c) \leq v_1 \leq 1$, then the denominator of the second logarithm changes sign. 

Putting all of this information together, one can see that (\ref{topj}) is easily expressed in terms of four double integrals:
\bea
&&\int_0^{1\over1+c} d v_2\int_0^{c\over 1+c} d v_1 {\bigg(\ln\bigg({N(c,c,-1,-1,v_1,v_2)\over c(1-v_1)-v_1}\bigg) + i \pi\bigg) \ln\bigg({N(c,c,-1,-1,v_1,v_2)\over 1-v_2-c v_2}\bigg) \over (1-2 v_1)(1-2 v_2)} \nonumber \\
&&+ \int_0^{c\over 1+c} d v_2 \int_{1-v_2 (1+c)\over(1+c)(1-2 v_2)}^1 d v_1 {\bigg(\ln\bigg({-N(c,c,-1,-1,v_1,v_2)\over -c(1-v_1)+v_1}\bigg) - i \pi\bigg) \bigg( \ln\bigg({-N(c,c,-1,-1,v_1,v_2)\over 1-v_2-c v_2}\bigg) - i \pi\bigg)\over (1-2 v_1)(1-2 v_2)} \nonumber \\
&&+ \int_{1\over 1+c}^1 d v_2 \int_{1-v_2 (1+c)\over(1+c)(1-2 v_2)}^{c \over 1+c} d v_1 {\bigg(\ln\bigg({N(c,c,-1,-1,v_1,v_2)\over c(1-v_1)-v_1}\bigg) + i \pi\bigg) \bigg( \ln\bigg({N(c,c,-1,-1,v_1,v_2)\over -1+v_2+c v_2}\bigg) + i \pi\bigg)\over (1-2 v_1)(1-2 v_2)}\nonumber \\
&&+ \int_{1\over 1+c}^1 d v_2 \int_{c \over 1+c}^1 d v_1 {\ln\bigg({N(c,c,-1,-1,v_1,v_2)\over -c(1-v_1)+v_1}\bigg)\bigg( \ln\bigg({N(c,c,-1,-1,v_1,v_2)\over -1+v_2+c v_2}\bigg) + i \pi\bigg)\over (1-2 v_1)(1-2 v_2)} \, .
\eea
The imaginary part of the above expression is easily extracted and is equal to 
\bea
&&i \pi \int_0^{1\over1+c} d v_2\int_0^{c\over 1+c} d v_1 { \ln\bigg({N(c,c,-1,-1,v_1,v_2)\over 1-v_2-c v_2}\bigg) \over (1-2 v_1)(1-2 v_2)} \nonumber \\
&&- i \pi \int_0^{c\over 1+c} d v_2 \int_{1-v_2 (1+c)\over(1+c)(1-2 v_2)}^1 d v_1 {\ln\bigg({-N(c,c,-1,-1,v_1,v_2)\over -c(1-v_1)+v_1}\bigg) + \ln\bigg({-N(c,c,-1,-1,v_1,v_2)\over 1-v_2-c v_2}\bigg) \over (1-2 v_1)(1-2 v_2)}\nonumber \\
&&+ i \pi \int_{1\over 1+c}^1 d v_2 \int_{1-v_2 (1+c)\over(1+c)(1-2 v_2)}^{c \over 1+c} d v_1 {\ln\bigg({N(c,c,-1,-1,v_1,v_2)\over c(1-v_1)-v_1}\bigg) + \ln\bigg({N(c,c,-1,-1,v_1,v_2)\over -1+v_2+c v_2}\bigg)\over (1-2 v_1)(1-2 v_2)}\nonumber \\
&&+ i \pi \int_{1\over 1+c}^1 d v_2 \int_{c \over 1+c}^1 d v_1 {\ln\bigg({N(c,c,-1,-1,v_1,v_2)\over -c(1-v_1)+v_1}\bigg)\over (1-2 v_1)(1-2 v_2)} \, .
\eea
This particular topology works well as an example because it is compact enough to be presented in detail and, nevertheless, it is one of the most intricate continuations that one must perform to reproduce the results given in this paper (working on the slice defined by Eq.~(\ref{slice1})).
\newpage

%%%%%%%%%%%%%%%%%%%%%%%%%%%%%%%%%%%%%%%%%%%%%%%%

\end{document}